\documentclass[12pt]{article}

\usepackage{subeqnarray}
\usepackage[mathscr]{eucal}
\usepackage{amsfonts}
\usepackage{amssymb}
\usepackage{amsthm}

\newcommand{\be}{\begin{equation}}
\newcommand{\ee}{\end{equation}}
\newcommand{\bea}{\begin{eqnarray}}
\newcommand{\eea}{\end{eqnarray}}

\begin{document}

\begin{titlepage}
\vspace*{1.5cm}

\renewcommand{\thefootnote}{\dag}
\begin{center}
{\LARGE\bf Nonrelativistic counterparts of twistors  and }

\vspace{0.5cm}

{\LARGE\bf  the realizations of Galilean conformal algebra}

\vspace{0.3cm}

{\LARGE\bf }

\vspace{1.5cm}
\renewcommand{\thefootnote}{\star}

{\large\bf S. Fedoruk}${}^{1}$,\,\,\,
{\large\bf P. Kosi\'{n}ski}${}^{2}$,\,\,\,{\large\bf J. Lukierski}${}^{3}$
,\,\,\,{\large\bf P. Ma\'{s}lanka}${}^{2}$\vspace{1cm}

${}^{1)}${\it Bogoliubov  Laboratory of Theoretical Physics, JINR,}\\
{\it 141980 Dubna, Moscow region, Russia} \\
\vspace{0.1cm}

{\tt fedoruk@theor.jinr.ru}, \\
\vspace{0.5cm}

${}^{2)}${\it Department of Theoretical Physics II, University of {\L}\'{o}d\'{z},}\\
{\it Pomorska 149/153,
90-236 {\L}\'{o}d\'{z}, Poland} \\
\vspace{0.1cm}

{\tt pkosinsk@uni.lodz.pl},\,\,\,\,{\tt pmaslan@uni.lodz.pl}\\
\vspace{0.5cm}

${}^{3)}${\it Institute for Theoretical Physics, University of Wroc{\l}aw,}\\
{\it pl.
Maxa Borna 9, 50-204 Wroc{\l}aw, Poland} \\
\vspace{0.1cm}

{\tt lukier@ift.uni.wroc.pl}\\
\vspace{0.9cm} \setcounter{footnote}{0}

05.04.2011

\end{center}
\vspace{0.2cm} \vskip 0.6truecm  \nopagebreak

\begin{abstract}
\noindent Using the notion of Galilean conformal algebra (GCA) in arbitrary space dimension $d$, we introduce
for $d{=}3$ quantized nonrelativistic counterpart of twistors as the spinorial representation of
${\rm SO}(2,1)\oplus{\rm  SO}(3)$ which is the maximal semisimple subalgebra of three--dimensional GCA.
The GC--covariant quantization of such nonrelativistic spinors,
which shall be called also Galilean twistors, is presented.
We consider for $d{=}3$ the general spinorial matrix realizations of GCA,
which are further promoted to quantum--mechanical operator representations, expressed as bilinears in quantized
Galilean twistors components. For arbitrary Hermitian quantum--mechanical Galilean twistor realizations we obtain
the result that the representations of GCA with positive--definite Hamiltonian do not exist.
For non--positive $H$ we construct for $N{\geq}2$ the Hermitian Galilean $N$--twistor realizations of GCA;
for $N{=}2$ such realization is provided explicitly.
\end{abstract}

\end{titlepage}

\newpage


\setcounter{equation}{0}
\section{Introduction}

\quad\, Since the introduction of $D{=}4$ twistors describing conformal ${\rm O}(4,2)\simeq{\rm SU}(2,2)$
spinors (see e.g.  \cite{Pen,PenMac}) as a basic alternative to the space-time coordinates there
were obtained many results  (see e.g. \cite{Wit78,Mad,Wit,Mason,BoMasSki}) using  relativistic twistorial framework.

The twistor $Z_A$ ($A=1,...,4$) carrying a particular $4{\times}4$ complex
matrix realization of $D{=}4$ conformal algebra spanned by
the generators $P_{\mu}$, $K_{\mu}$,
$M_{\mu\nu}=(M_{i}{=}\frac{1}{2}\, \epsilon_{ijk} M_{jk}, N_{i}{=}M_{i0})$, $D$
($\mu=0,1,2,3$; $i=1,2,3$) (see e.g. \cite{Wells,JakOdz})
\begin{equation}\label{alg-conf}
\begin{array}{c}
P_{\mu}=
\left(
  \begin{array}{cc}
    0 & 0 \\
    \sigma_{\mu} & 0 \\
  \end{array}
\right),\qquad
K_{\mu}=
\left(
  \begin{array}{cc}
    0 & \sigma_{\mu} \\
    0 & 0 \\
  \end{array}
\right), \\
M_{i}=\frac12
\left(
  \begin{array}{cc}
    \sigma_{i} & 0 \\
    0 & \sigma_{i} \\
  \end{array}
\right),\qquad
N_{i}=\frac12
\left(
  \begin{array}{cc}
    \sigma_{i} & 0 \\
    0 & -\sigma_{i} \\
  \end{array}
\right),\qquad
D={\textstyle\frac{i}{2}}
\left(
  \begin{array}{cc}
    \sigma_{0} & 0 \\
    0 & -\sigma_{0} \\
  \end{array}
\right)
\end{array}
\end{equation}
can be described by two $D{=}4$ Weyl spinors
\begin{equation}\label{tw-two-sp}
Z_{A}=
\left(
  \begin{array}{c}
    \lambda_\alpha \\
    \mu^{\dot\alpha} \\
  \end{array}
\right),\qquad
\alpha =1,2\,.
\end{equation}
If we perform the nonrelativistic contraction of relativistic conformal algebra
to GCA (see (\ref{coset1})-(\ref{contr2})),
the finite--dimensional matrix realizations of the generators $P_i$, $N_i$, $K_i$ after contraction in the Abelian
limit are nullified, and nonvanishing remain only the following noncontracted ones
\begin{equation}\label{alg-contract}
M_{i}={\textstyle\frac{1}{2}}
\left(
  \begin{array}{cc}
    \sigma_{i} & 0 \\
    0 & \sigma_{i} \\
  \end{array}
\right),\qquad
P_{0}=
\left(
  \begin{array}{cc}
    0 & 0 \\
    \sigma_{0} & 0 \\
  \end{array}
\right),\quad
K_{0}=
\left(
  \begin{array}{cc}
    0 & \sigma_{0} \\
    0 & 0 \\
  \end{array}
\right),\quad
D={\textstyle\frac{i}{2}}
\left(
  \begin{array}{cc}
    \sigma_{0} & 0 \\
    0 & -\sigma_{0} \\
  \end{array}
\right)
.
\end{equation}
Introducing notation
\begin{equation}\label{nonrel-rel}
t_{\alpha,\,1}=\lambda_{\alpha},\qquad
t_{\alpha,\,2}=\mu^{\dot\alpha}
\end{equation}
we obtain that the $2{\times}2$ matrix $t_{\alpha,\,a}$ ($a=1,2$) becomes a spinorial
representation of ${\rm SU}(2){\otimes}{\rm SU}(1,1)\simeq{\rm O}(3){\otimes}{\rm O}(2,1)$,
with generators $M_i\in o(3)$ and
$(P_{0},K_{0},D)\in o(2,1)$ which is the maximal semisimple
non-Abelian subalgebra of GCA. In such a way we shall speak further about the nonrelativistic limit
of relativistic twistor, which will be called as nonrelativistic counterpart of twistors
or Galilean twistors.\footnote{
Twistors in $D{=}4$ were also defined as the coordinates of twistor bundle
over four--dimensional space--time ${\mathscr{M}}$ with the fibre described by all complex structures on ${\mathscr{M}}$.
However, because the nonrelativistic space--time splits into
one--dimensional time parameters manifold and three--dimensional
nonrelativistic space, one cannot introduce in such odd--dimensional space a
complex structure, and the nonrelativistic version
of the definition given above cannot be applied. Similarly,
if we define relativistic twistors via cosets, e.g. as
projective twistors parametrizing the coset
$\textstyle{\frac{{\rm SU}(2,2)}{{\rm S}\left({\rm U}(1)\times{\rm U}(1,2)\right)}}$,
the contractions providing the Galilean conformal group from
${\rm SU}(2,2)$ (see (\ref{contr1}) and (\ref{contr2})) do not provide a sensible
nonrelativistic limit.
}

Two basic ingredients of relativistic twistors approach are \cite{PenRin,Hug}
\begin{description}
  \item[i)] The expression of the Poincar\'{e} and conformal algebra generators
  as bilinears in terms of quantized twistor coordinates.
  In particular, in relativistic $N$--twistor space
  $Z^{(k)}_A=(\lambda_{\alpha}^{(k)}, \mu^{\dot\alpha(k)})$ ($k=1,...,N$)
  the four-momentum generators $P_\mu=\frac12\,P^{\alpha\dot\beta}(\sigma_\mu)_{\alpha\dot\beta}$ are represented by the formula
  \begin{equation}\label{mom-tw}
  P_{\alpha\dot\beta} = \sum_k\lambda_{\alpha}^{(k)}\bar\lambda_{\dot\beta}^{(k)}\,, \qquad
  \lambda^{(k)}_{\dot{\beta}} = (\lambda^{(k)}_\beta)^{+}
  \end{equation}
  with positive--definite energy component $P_0$.
  The relation  (\ref{mom-tw}) is consistently extended to
  Poincar\'{e} and conformal algebra.\footnote{The extension to superconformal algebra
  has been also proposed \cite{Fer}.}
  \item[ii)] The second Lorentz spinors $\mu^{\dot\alpha(k)}$ defining twistors satisfy
  Cartan--Penrose incidence relation
  \begin{equation}\label{inci-tw}
  \mu^{\dot\alpha(k)}=X^{\dot\alpha\beta} \lambda_{\beta}^{(k)}
  \end{equation}
  linking complexified space--time with twistor coordinates.
  For $N{\geq} 2$ one can express space--time coordinates $X^{\dot\alpha\beta}$
  as composite in terms of twistorial components.
\end{description}

The aim of this Letter is to study
the spinorial realizations of Galilean conformal algebra (GCA) and consider their applicability to the description of
nonrelativistic dynamics. We shall concentrate on the possibility of construction
of the quantum--mechanical realizations of GCA in terms of Galilean twistors with positive--definite
Hamiltonian $H$.

In order to introduce nonrelativistic conformal symmetry we shall define in Sect.~2 the conformal
limit of $D{=}4$ conformal algebra which describes GCA \cite{LSZ06,LSZ07,BaGo,DuHor}.\footnote{
The general mathematical considerations providing as special case GCA were given in \cite{NOR-M}.}
We stress that GCA is different from the Schr\"{o}dinger algebra, which was also
named as nonrelativistic conformal algebra  (see e.g.  \cite{Hag,DuHor}).
Further, in Sect. 3, we consider the realizations of GCA in terms of Galilean twistors $t_{\alpha,\,a}$
($\alpha=1,2$; $a=1,2$) which, we recall, are the fundamental spinor representation of the semisimple part of GC symmetry
${\rm SU}(2){\otimes}{\rm SU}(1,1)\simeq{\rm O}(3){\otimes}{\rm O}(2,1)$. We stress that, contrary to the relativistic case,
the nonrelativistic GCA is not semisimple, and in $d$ space dimensions
it can written as the following  $\frac{(d+2)(d+3)}{2}$--dimensional semidirect sum
\begin{equation}\label{GCA-Str}
{\mathscr{C}}^{(d)}=\Big( o(2,1)\oplus o(d) \Big) \ltimes {\mathscr{A}}^{(3d)}
\end{equation}
where $o(2,1)$ describes the conformal symmetries on the world line,
$o(d)$ generates the space rotations, and ${\mathscr{A}}^{(3d)}$ describes $3d$--dimensional Abelian subalgebra
of space translations, Galilean boosts and nonrelativistic constant accelerations.
The Galilean $\frac{(d+1)(d+2)}{2}$--dimensional algebra in $d$ space dimensions
\begin{equation}\label{GA-Str}
{\mathscr{G}}^{(d)}=\Big( o(1,1)\oplus o(d) \Big) \ltimes \tilde{\mathscr{A}}^{(2d)}
\end{equation}
is a subalgebra of ${\mathscr{C}}^{(d)}$, ${\mathscr{G}}^{(d)}\subset {\mathscr{C}}^{(d)}$,
where in (\ref{GA-Str}) $o(1,1)$ generator is the Hamiltonian $H$, and $\tilde{\mathscr{A}}^{(2d)}$
describes space translations and Galilean boosts.

In order to preserve ${\rm SO}(2,1)\oplus{\rm  SO}(3)$ nonrelativistic conformal
invariance\footnote{For simplicity further we shall consider the physical case $d{=}3$.}
 we postulate for
 Galilean twistors the following CCR
\begin{equation}\label{CCR-tw1}
\left[t_{\alpha,\,a}^{(k)},t_{\dot\beta,\,\dot b}^{+(l)} \right]=\delta^{kl}\delta_{\alpha\dot\beta}\,\omega_{a\dot b}
\end{equation}
where $\delta_{\alpha\dot\beta}$ describes the standard unitary metric in $\mathbb{C}^2$
and $\omega_{a\dot{b}}$ is a $2{\times}2$ traceless Hermitian matrix describing ${\rm SU}(1,1)$
metric.\footnote{The indices of ${\rm SU}(2)$ spinors can be raised and lowered by the rule
$(\psi_{\alpha})^+{=}\,\bar\psi_{\dot\alpha}{=}\,\bar\psi{}^{\alpha}$,
$\phi_{\alpha}{=}\,\phi^{\dot\alpha}$ what leads to equivalent two ways
$\bar\psi_{\dot\alpha}\phi_{\alpha}{\equiv}\,\bar\psi^{\alpha}\phi_{\alpha}$
of describing ${\rm SU}(2)$--invariant scalar product.
For ${\rm SU}(1,1)$ spinors we have the following notations $\psi_{a}{=}\,\omega_{a\dot b}\psi^{\dot b}$,
$\phi_{\dot a}{=}\,\phi^{b}\omega_{b\dot a}$.}
It should be observed that
 vanishing trace condition implies indefinite metric what can be seen explicitly from (\ref{CCR-tw1})
e.g. by choosing $\omega=\sigma_3$.
If we redefine $t_{\alpha,\,1}^{(k)}=u_{\alpha,\,1}^{(k)}$, $t_{\alpha,\,2}^{(k)}= (u_{\alpha,\,2}^{(k)} )^+$, the relations (\ref{CCR-tw1})
equivalently can be written as follows
\begin{equation}\label{CCR-tw2}
\left[u_{\alpha,\,a}^{(k)},u_{\dot\beta,\,\dot b}^{+(l)}
\right]=\delta^{kl}\delta_{\alpha\dot\beta}\delta_{a\dot b}\,.
\end{equation}
Relations (\ref{CCR-tw2}) lead to the Hilbert space realization with positive metric but with broken
${\rm  SU}(1,1)\simeq{\rm SO}(2,1)$ subgroup of GCA.
If we use the assignment (\ref{nonrel-rel})
and choose $\omega_{a\dot b}$ given by $\sigma_1$,
the relations (\ref{CCR-tw1}) describe  the Penrose
${\rm  SU}(2,2)$--covariant twistor quantization
\begin{equation}\label{CCR-tw3}
\left[\lambda_{\alpha}^{(k)},\bar\mu^{\beta (l)} \right]=\delta^{kl}\delta_{\alpha}^{\beta}\,,\qquad
\left[\mu^{\dot\beta (l)},\bar\lambda_{\dot\alpha}^{(k)} \right]=\delta^{kl}\delta_{\dot\alpha}^{\dot\beta}\,.
\end{equation}
 We see therefore that the indefinite metric quantization
is  present as well  in the relativistic covariant twistor quantization.
It can be added that the alternative noncovariant twistor quantization, corresponding to the choice (\ref{CCR-tw2})
of Galilean twistor oscillators, has been used in relativistic case by Woodhouse \cite{Wood}.

In Sect.~3 we shall introduce the $4N{\times}4N$--dimensional complex reducible realization
of the GCA describing the nonrelativistic conformal transformations of the
multiplet of $N$ Galilean ${\rm SO}(3)\otimes{\rm O}{(2,1)}$ spinors.
Because contrary to the relativistic case it is not possible to introduce the Galilean twistor realization of GCA if
$N{=}1$, we shall consider $N\geq{2}$.

We describe such matrix realizations as factorized
into products of three matrices.
To describe the matrix realization $C_R$ of GCA generators $\hat C_R$ = ($P_i$, $H$, $M_i$, $B_i$, $D$, $K$, $K_i)$
(see Sect.~2) we start with the following general $4N{\times}4N$ matrices
\begin{equation}\label{tens-pr1}
\left(C_R \right)_{\alpha,\,a,\,k}{}^{\beta,\, b,\,l}=
\sum\,  (\sigma_{\mu})_{\alpha}{}^{\beta}
\otimes
(\sigma_\nu)_{a}{}^{{b}}\otimes ({\mathscr{A}}_R^{\mu\nu})_{k}{}^{l}
\end{equation}
where $\sigma_\mu = (\sigma_i,1)$, $i=1,2,3$
describe the scalar and spinorial $su(2)$ algebra realizations,
$\rho_\mu =(\rho_r,1)$, $r=0,1,2$, $\rho_0{=}\,\sigma_1$, $\rho_1{=}\,i\sigma_2$, $\rho_2{=}\,i\sigma_3$
provide the scalar and spinorial  realization of $su(1,1)$ algebra and
${\mathscr{A}}^{\mu\nu}_R$ is the set of $N{\times}N$
matrices specified after the substitution of (\ref{tens-pr1}) in place of $\hat C_R$ into GCA.
Having the quantized twistor oscillators (\ref{CCR-tw1}) and matrix realizations $C_R$ of
the generators of GCA
one can introduce the quantum--mechanical $N$-twistor realization
of GCA as follows\footnote{We recall that the  correspondence with relativistic twistor realization (\ref{CCR-tw3}) requires $\omega=\sigma_1$.}
\begin{equation}\label{real-op1}
\hat C_R=\bar t_{\dot\alpha,\,\dot a}^{\,\,(k)}\, \delta^{\dot\alpha\alpha}
\omega^{\dot a a}\left(C_R \right)_{\alpha,\,a,\,k}{}^{\beta,\, b,\,l}\,
t_{\beta,\,b}^{(l)}
\end{equation}
where $\bar t_{\dot\alpha,\,\dot a}^{\,\,(k)}=t_{\dot\alpha,\,\dot a}^{+(k)}$ and  $\omega^{\dot a c}\omega_{c\dot b}=\delta^{\dot a}_{\dot b}$.
We see that the formula
(\ref{real-op1}) promotes the ``classical'' matrix representation (\ref{tens-pr1}) of GCA to quantum--mechanical operator realization.

In our Letter we shall investigate the possibility of constructing Hermitian GCA realization with
positive--definite Hamiltonian $H$ in agreement with the postulates of QM.
Firstly in Sect.~3 we shall consider the case $N{=}2$, and further we extend the discussion to arbitrary $N$.
Unfortunately the result is negative -- for the choice of any $N$ it does not exist a matrix realization (\ref{tens-pr1})
which leads to formula (\ref{real-op1}) describing positive--definite Hamiltonian.
It will be shown however in Sect.~4 that in nonrelativistic case the Hermitian twistorial operator realizations with indefinite Hamiltonians
do exist for any $N\geq{2}$.
We see therefore that the applicability of proposed Galilean twistors realization of GCA is limited
to the description of somewhat exotic models with indefinite Hamiltonians.

\setcounter{equation}{0}
\section{Galilean conformal algebra and Galilean twistors}

\quad\, The Galilean conformal algebra can be obtained by a contraction of $D{=}d+1$--dimensional
relativistic conformal algebra $o(D,2)$ with the Poincar\'{e} generators
$P_\mu=(P_0, P_i)$, $M_{\mu\nu}=(M_{ij},M_{i0})$ ($i,j=1,...,d$, $\mu,\nu=0,1,...,d$),
dilatation generator $D$ and $K_\mu=(K_0, K_i)$ generating special conformal transformations.
The contraction procedure is not unique. If we wish to identify the contraction parameter
as a light velocity $c$ the ``physical'' nonrelativistic contraction is defined by the following rescaling
\cite{LSZ06}
\begin{equation}\label{contr1}
P_0=\frac{H}{c}  \, , \qquad  M_{i 0}=c\, B_i \, , \qquad K_0 = c\, K \, ,  \qquad   K_i =
c^{2} F_i
\end{equation}
($P_i$, $M_{ij}$ and $D$ remaining unchanged). After performing the nonrelativistic
contraction limit $c\to \infty$ one obtains the following $\frac12\,(D+1)(D+2)$--dimensional
Galilean conformal algebra ${\mathscr{C}}^{(d)}$ \cite{LSZ06, LSZ07,BaGo}
\begin{equation}\label{H-A}
 [H,P_i]=0 \, ,  \quad   [H,B_i]= iP_i \, , \quad  [H,F_i]=2iB_i\, ,
\end{equation}
\begin{equation}\label{K-A}
[K, P_i]=-2iB_i \, , \quad  [K,B_i]=-iF_i\, , \quad  [K, F_i]=0 \, ,
\end{equation}
\begin{equation}\label{D-A}
[D,P_r]=- iP_i \, , \quad  [D,B_i]=0 \, , \quad   [D,F_i]=iF_i \, ,
\end{equation}
and
\begin{equation}\label{su11}
[D,H] = - iH \, , \quad [K,H]=-2iD \, , \quad [D,K]=iK\,,
\end{equation}
where the subalgebra $(P_i,B_i,F_i)$ describes the generators of maximal Abelian
subgroup and the subalgebra $(H,D,K)$ span the  $o(2,1)$ algebra.
The algebra of  rotations $ o(d)$ is described by the commutators
\begin{equation}\label{sod}
 [M_{ij}, M_{k l}]= i\left(\delta_{ik} \,M_{j l} - \delta_{il}\, M_{sk} + \delta_{jl}\,
M_{ik} - \delta_{jk} M_{il}\right)\,,
\end{equation}
and
\begin{equation}\label{M-A}
\!\!\!\!\!\!\!\!\!\! [M_{ij},P_{k}]{=}\,i\left(\delta_{ik} P_{k} - \delta_{jk} P_{l}\right)  ,\quad
[M_{ij},F_{k}]{=}\,i\left(\delta_{ik} F_{k} - \delta_{jk} F_{l}\right)  ,\quad
[M_{ij},B_{k}]{=}\,i\left(\delta_{ik} B_{k} - \delta_{jk} B_{l}\right)  ,
\end{equation}
\begin{equation}\label{M-T}
[M_{ij}, H]=[M_{ij}, K]=[M_{ij}, D]=0\,.
\end{equation}
The generators ($P_i,B_i, H, M_{ij}$) define the Galilean algebra ${\mathscr{C}}^{(d)}$, where $B_i$ are the
Galilean boosts and $H$, the nonrelativistic energy operator, generates the Galilean time
translations. We see that one can treat the Galilean conformal algebra as the
 result of adding the generators $D$ (dilatations), $K$
(expansions)  and
 $F_i$ (constant accelerations)  to the Galilean algebra.

The Galilean conformal algebra  ${\mathscr{C}}^{(d)}$ can be described as the semidirect product
(\ref{GCA-Str}). Such a structure indicates that GCA can be also obtained as the contraction of the following
coset of ${\rm O}(d+1,2)$
\begin{equation}\label{coset1}
{\mathscr{K}}=\frac{{\rm O}(d+1,2)}{{\rm O}(2,1)\otimes{\rm O}(d)}\,.
\end{equation}
In such a case the rescaling of relativistic conformal generators is introduced in the following way
\begin{equation}\label{contr2}
P_i=\kappa\, P_i^{NR}  \, , \qquad  M_{i 0}=\kappa\, B_i \, , \qquad K_i =\kappa\,  F_i
\end{equation}
and the generators $P_0$, $K_0$, $M_{ij}$ and $D$ are not changed.
In such a framework
 we should treat the
  $o(d+1,2)$
   generators
  and the parameter $\kappa$ as dimensionless. It can be shown that
the alternative ``formal'' nonrelativistic rescaling (\ref{contr2}) provides as well in
the limit $\kappa\to\infty$ the GCA given by  (\ref{H-A})--(\ref{M-T}) \cite{GGK}.
 The lack of uniqueness of the contraction procedure providing GCA follows from the possible
scaling automorphisms of relativistic conformal algebra, represented e.g. by the following mapping
($X_A$ are the conformal algebra generators)
\begin{equation}\label{auto}
X^{\xi}_A=e^{\xi D}\, X_A\, e^{-\xi D}\,.
\end{equation}
It can be shown that by suitable choice of $\xi$ the rescaling  (\ref{contr2}) of the generators
$X_A$ reproduce the rescaling (\ref{contr1}) for the generators $X^{\xi}_A$.

As follows from (\ref{GCA-Str}), if $d{=}3$, the maximal semisimple subgroup of GC group
is ${\rm O}(2,1)\otimes{\rm O}(3)$, and Galilean twistors
$t_{r,\,a}$ describe the fundamental representation of its spinor covering
$\overline{{\rm O}(2,1)}{\otimes}\overline{{\rm O}(3)}={\rm SU}(1,1){\otimes}{\rm SU}(2)$.
Considering twistor as $2{\times}2$ matrix $\mathbb{T}= (t_{\alpha,a})$ one obtains the following
GC transformations
\begin{equation}\label{tw-trans}
\mathbb{T}^\prime=A\,\mathbb{T}\,B\,,\qquad A\in {\rm SU}(2)\,,\qquad B\in {\rm SU}(1,1)\,.
\end{equation}
The basis of $o(2,1)\oplus o(3)$ Lie algebra is described by the generators of relativistic conformal
algebra which were not contracted in the contraction procedure $\kappa\to\infty$ (see  (\ref{contr2})),
namely
\begin{equation}\label{non-contr}
\begin{array}{rcl}
o(2,1)&:&\qquad P_0=H\,,\quad K_0=K\,,\quad D\,, \\
o(3)&:&\qquad M_{i}={\textstyle\frac12}\,\epsilon_{ijk}M_{jk}\,.
\end{array}
\end{equation}
Introducing standard ${\rm O}(2.1)$ basis
\begin{equation}\label{T-def}
R_0={\textstyle\frac12}\,(K+H)\,,\qquad R_1={\textstyle\frac12}\,(K-H)\,,\qquad R_2=D
\end{equation}
one gets two commuting three--dimensional Lie algebras
 ($r,s,t=0,1,2$; $i,j,k=1,2,3$)
\begin{eqnarray}\label{alg-non-contr}
o(2,1)&:&\qquad\left[R_{r}, R_{s}\right]=i\,\epsilon_{rst} \,R^{t} \,,
\\
\label{alg-non-contr-b}
o(3)&:&\qquad \left[M_{i}, M_{j}\right]=i\,\epsilon_{ijk}\,M_{k}
\end{eqnarray}
where $\epsilon_{123}=\epsilon_{012}=+1$, $R^{r}=g^{rs}R_{s}$,
$g_{rs}{=}{\rm diag}(-++)$.
The contracted part of the relativistic conformal algebra $o(4,2)$ describes the Abelian generators
$P_i$, $B_i$, $F_i$.

 If we introduce new basis of Abelian subalgebra of GCA
 \begin{equation}\label{yyy2.17}
 {\mathscr{A}}_{i,0} = {\textstyle\frac12}\,(F_i + P_i)\,,
 \qquad
 {\mathscr{A}}_{i,1} = {\textstyle\frac12}\,(F_i - P_i)\,,
 \qquad
 {\mathscr{A}}_{i,2} = B_i
 \end{equation}
 we obtain the following $o(2,1)\oplus o(3)$ extension of (\ref{M-A})
 \begin{equation}\label{yyy2.18}
 [M_i, {\mathscr{A}}_{j,r}] = i \, \varepsilon_{ijk}
 {\mathscr{A}}_{k,r} \,,
 \qquad
 [R_r, {\mathscr{A}}_{i,s}] = i\, \varepsilon_{rs}{}^{t}{\mathscr{A}}_{i,t}
 \end{equation}

The last considerations (formulae  (\ref{tw-trans})--(\ref{yyy2.18})) can be easily extended to arbitrary $d$.

\setcounter{equation}{0}
\section{Multitwistor realizations of GCA}

\quad\, We shall consider extended ($N{\geq}2$) Galilean twistor space
because the nontrivial $4{\times}4$ matrix realizations of all generators of GCA on simple $N{=}1$ Galilean
twistor space do not exist. In order to show it let us note that
the generators of the direct product of $su(2)$ and $su(1,1)$ generators in $4{\times}4$ space are
\begin{equation}\label{4-MR}
M_i={\textstyle\frac12}\,\sigma_i \otimes \mathbf{1}_2\,,\qquad
R_{r}=\mathbf{1}_2 \otimes {\textstyle\frac12}\,\rho_{r}
\end{equation}
where $\rho_0{=}\,\sigma_1$, $\rho_1{=}\,i\sigma_2$, $\rho_2{=}\,i\sigma_3$.
Then, the only nine generators which transform as the product of vectorial representations of $su(2)$ and $su(1,1)$
are (see (\ref{yyy2.18}))
\begin{equation}\label{Matr-rel-A1}
{\mathscr{B}}_{i,r}= \sigma_i \otimes \rho_{r} \,.
\end{equation}
The generators (\ref{Matr-rel-A1}) are the only candidates to describe the  $4{\times}4$ matrix realization
of generators ${\mathscr{A}}_{i,r}$, but they obviously do not form Abelian subalgebra. In fact,
15 generators (\ref{4-MR}) and (\ref{Matr-rel-A1}) form $su(2,2)$ algebra
and using the construction (\ref{real-op1}) and the identification (\ref{nonrel-rel})
we obtain standard $N{=}1$ twistor realization of relativistic conformal algebra.

\subsection{The realizations with  positive-definite Hamiltonian $H$}

\quad\, Let us firstly consider Galilean twistor realizations of GCA for $N{=}2$,
with all three factors in tensor product (\ref{tens-pr1}) described by $2{\times}2$
matrices. Such representations provides four doublets of ${\rm SU}(2)$.
We  choose the following spinorial realization
of the generators of  $su(2)$ acting on four nonrelativistic spinors
\begin{equation}\label{matrix-M}
M_i={\textstyle\frac12}\,\sigma_i \otimes \mathbf{1}_2 \otimes \mathbf{1}_2
\end{equation}
and the spinorial generators of ${\rm SU}(1,1)$ (in consistency with $[R_a, M_i]=0$)
\begin{equation}\label{matrix-T}
R_{r}=\mathbf{1}_2 \otimes {\textstyle\frac12}\,\rho_{r} \otimes \mathbf{1}_2
\end{equation}
where $\rho_0{=}\,\sigma_1$, $\rho_1{=}\,i\sigma_2$, $\rho_2{=}\,i\sigma_3$.
 From  (\ref{T-def}) we obtain that
 ($\sigma_{\pm}{=}\,{\textstyle\frac{1}{2}}(\sigma_1{\pm}i\sigma_2)$)
\begin{subeqnarray}
\label{matrix-T-conc}
H&{=}& \mathbf{1}_2 \otimes
\sigma_{-}
 \otimes \mathbf{1}_2\,,
 \\
K&{=}&  \mathbf{1}_2 \otimes
\sigma_{+}
\otimes \mathbf{1}_2\,,
\\
D&{=}&  \mathbf{1}_2 \otimes {\textstyle\frac{i}{2}}\,\sigma_3 \otimes \mathbf{1}_2
\end{subeqnarray}

 The covariance relations (\ref{yyy2.18})
imply that one can choose
\begin{equation}\label{Matr-A1}
{\mathscr{A}}_{i,r}= \sigma_i \otimes \rho_{r} \otimes {\mathscr{C}}
\end{equation}
where
 the complex   $2{\times}2$ matrix
\begin{equation}\label{Matr-A2x2}
{\mathscr{C}}= c_0 + \vec{c}\, \vec{\sigma}
\end{equation}
is unique for all nine generators ${\mathscr{A}}_{i,r}$.

The $2{\times}2$ matrix ${\mathscr{C}}$ is specified by the   Abelian nature of ${\mathscr{A}}_{\,i\, , r}$,
\begin{equation}\label{A-A-com1}
[{\mathscr{A}}_{\,i\, , r},{\mathscr{A}}_{\,j \,, s}]= 0\,.
\end{equation}
The relations  (\ref{A-A-com1}) requires that matrix ${\mathscr{C}}$ is nilpotent, i.e. ${\mathscr{C}}^2=0$,
or more explicitly
\begin{equation}\label{alpha-conds}
{\mathscr{C}}^2=0\qquad\quad \Rightarrow \quad\qquad  c_{0}=0, \quad \vec{c}{}^{\,\, 2}=0\,.
\end{equation}

Solving   (\ref{alpha-conds}) we obtain two two--parameter families
\begin{equation}\label{Matr-A2x2-2par}
{\mathscr{C}}_1= \left(\begin{array}{cc}
                      a & b \\
                      -a^2/b & -a \\
                      \end{array}
                      \right)\,,
\qquad\qquad{\mathscr{C}}_2= \left(\begin{array}{cc}
                      a & -a^2/b \\
                      b & -a \\
                      \end{array}
                      \right)
\end{equation}
where $b\neq0$, whereas $a$ is arbitrary.
One can prove by considering the equivalent realizations of $o(2,1)$
\begin{equation}\label{yyy3.9}
\widetilde{\rho}_r = V \, \rho_r \, V^{-1} \,,
\end{equation}
where $V$ is an arbitrary invertible $2{\times}2$ matrix,
that all nontrivial choices of $a$, $b$ are equivalent.
We can take the special case
$c_0 = c_3 =0$, $c_1 = ic_2= \textstyle{\frac{1}{2}}$ corresponding to
$b=1$, $a=0$, i.e.
\begin{equation}\label{Matr-A2x2-fin}
{\mathscr{C}}= \left(\begin{array}{cc}
                      0 & 1 \\
                      0 & 0 \\
                      \end{array}
                      \right)\,.
\end{equation}
All other $2{\times}2$ nilpotent matrices (\ref{Matr-A2x2-2par}) can be obtained by similarity
transformation (\ref{yyy3.9}).

Let us apply now the formula (\ref{real-op1}), with the choice $\omega=\sigma_1$ and ${\mathscr{C}}$
given by (\ref{Matr-A2x2-fin})
in order to obtain from  (\ref{matrix-M}), (\ref{matrix-T}) and (\ref{Matr-A1}) the formulae for the generators
$\hat M_i$, $\hat H$, $\hat K$, $\hat D$, $\hat P_i$, $\hat B_i$ and $\hat K_i$. One gets,
in case of $\hat H$ in accordance with positive--definite formula (\ref{mom-tw})
(see also footnote 4)
\begin{equation}\label{M-op}
\hat M_i= {\textstyle\frac{1}{2}} \sum_{k=1}^2 \left(\bar t_{\dot\alpha ,1}^{(k)}(\sigma_i)_{\alpha}{}^{\beta}t_{\beta ,2}^{(k)}-
\bar t_{\dot\alpha ,2}^{(k)}(\sigma_i)_{\alpha}{}^{\beta}t_{\beta ,1}^{(k)}\right)
\end{equation}
\begin{equation}\label{M-R-op}
\hat H= \sum_{k=1}^2\,\bar t_{\dot{\alpha}, 1}^{(k)}\, t_{\alpha, 1}^{(k)}\,, \qquad
\hat K= \sum_{k=1}^2\,\bar t_{\dot{\alpha}, 2}^{(k)}\, t_{\alpha, 2}^{(k)}\,, \qquad
\hat D= {\textstyle\frac{i}{2}}\sum_{k=1}^2 \left(\bar t_{\dot\alpha ,2}^{(k)}\,t_{\alpha ,1}^{(k)}-
\bar t_{\dot\alpha ,1}^{(k)}\,t_{\alpha ,2}^{(k)}\right)
\end{equation}
and (see also  (\ref{Matr-A1}))
\begin{equation}\label{A-un-op}
\!\!\!\!
\hat P_i=\bar t_{\dot\alpha , 1}^{(1)}
\,(\sigma_i)_{\alpha}{}^{\beta}\,
t_{{\beta},1}^{(2)}
 \,,\quad\,\,\,\,
\hat F_i=\bar t_{\dot\alpha , 2}^{(1)}
\,(\sigma_i)_{\alpha}{}^{\beta}\,
t_{{\beta},2}^{(2)}
 \,, \quad\,\,\,\,
\hat B_i= {\textstyle\frac{i}{2}}
\left(\bar t_{\dot\alpha ,2}^{(1)}\,(\sigma_i)_{\alpha}{}^{\beta}\,t_{\beta ,1}^{(2)}-
\bar t_{\dot\alpha ,1}^{(1)}\,(\sigma_i)_{\alpha}{}^{\beta}\,t_{\beta ,2}^{(2)}\right)
\!.
\end{equation}

All the operator realizations (\ref{Matr-A1}) of the generators $P_i, B_i$ and $F_i$ are  not Hermitian.
This property remains valid if we insert in the relations (\ref{real-op1}) the matrix  realization (\ref{Matr-A1})
 with general nilpotent matrix (see (\ref{alpha-conds})).
If we assume the Hermiticity of (\ref{Matr-A1}), i.e. ${\mathscr{C}}={\mathscr{C}}^{+} $
we obtain from (\ref{Matr-A2x2}) and (\ref{alpha-conds})
\begin{equation}\label{yyy3.13}
{\mathscr{C}}\, {\mathscr{C}}^{+} = 0 \qquad \Rightarrow \qquad
|\,\vec{c}\,|^{\, 2} = 0 \qquad \Rightarrow \qquad  \quad c_i =0
\end{equation}

The argument about nonexistence of the Hermitian representation of GCA can be extended to arbitrary
multiplet of Galilean twistors ($k=1,...,N$). If we assume positive $H$, i.e.
choose unique formula
\begin{equation} \label{H-N}
\hat H=\sum_{k=1}^N\, \bar t_{\dot\alpha, 1}^{(k)} \, t_{\alpha,1}^{(k)}
\end{equation}
we should
 choose the following $4N{\times}4N$ matrix realization of H
\begin{equation}\label{H-N-op}
H= \mathbf{1}_2 \otimes
\sigma_{-} \otimes \mathbf{1}_N
\end{equation}
implying that
\begin{equation}\label{KD-N-op}
\begin{array}{rcl}
K&=&  \mathbf{1}_2 \otimes
\sigma_{+} \otimes \mathbf{1}_N\,,\\
D&=&  \mathbf{1}_2 \otimes {\textstyle\frac{i}{2}}\,\sigma_3 \otimes \mathbf{1}_N\,.
\end{array}
\end{equation}
Further
\begin{equation}\label{matrix-M-N}
M_i={\textstyle\frac12}\,\sigma_i \otimes \mathbf{1}_2 \otimes \mathbf{1}_N
\end{equation}
\begin{equation}\label{matrix-T-N}
R_{r}=\mathbf{1}_2 \otimes {\textstyle\frac12}\,\rho_{r} \otimes \mathbf{1}_N
\end{equation}
and ${\rm O}(3)\otimes {\rm O}(2,1)$ covariance relations
(\ref{Matr-A2x2})
imply that for the Abelian GCA  generators
one obtains the compact matrix  relation
\begin{equation}\label{Matr-A1-N}
{\mathscr{A}}_{\,i\,, r}= \sigma_i \otimes \rho_{r} \otimes {\mathscr{C}}^{(N)}
\end{equation}
where the  commutativity of the generators ${\mathscr{A}}$
implies nilpotency condition for $N{\times}N$ matrix ${\mathscr{C}}^{(N)}$
\begin{equation}\label{nil-A-N}
\left({\mathscr{C}}^{(N)}\right)^2=0\,.
\end{equation}
From (\ref{Matr-A1-N}) and (\ref{real-op1}) for any choice of the $2{\times}2$ spinorial $O(2,1)$
metric $\omega^{ab}$ follows that the Hermiticity condition implies
\begin{equation}\label{Her-A-N}
\left({\mathscr{C}}^{(N)}\right)^+={\mathscr{C}}^{(N)}
\end{equation}
what together with conditions (\ref{nil-A-N}) implies that ${\mathscr{C}}^{(N)}\left({\mathscr{C}}^{(N)}\right)^+=0$,
i.e. ${\mathscr{C}}^{(N)}=0$.
Indeed, by the similarity transformation any Hermitian matrix
can be written as the diagonal one with real eigenvalues;  the condition (\ref{nil-A-N})
implies that these eigenvalues are all equal to zero.

\subsection{The realizations with  indefinite Hamiltonian $H$}

\quad\, For $N=2$ indefinite  $H$ means the choice (compare with (\ref{matrix-T-conc}.a))
\begin{equation}\label{yyy3.22}
H =  \mathbf{1}_2 \otimes \sigma_{-} \otimes \sigma_3
\end{equation}
and the generators of  $su(2)$ as in  (\ref{matrix-M}):
\begin{equation}\label{matrix-M1}
M_i={\textstyle\frac12}\,\sigma_i \otimes \mathbf{1}_2 \otimes \mathbf{1}_2\,.
\end{equation}
The remaining generators of $o(2,1)$ subalgebra follow uniquely
\begin{eqnarray}\label{yyy3.23}
K & = &   \mathbf{1}_2 \otimes \sigma_+ \otimes \sigma_3 \,,
\cr
D & = &  \mathbf{1}_2 \otimes {\textstyle\frac{i}{2}} \, \sigma_3 \otimes 1_2 \,.
\end{eqnarray}
One can comment that the existence of doublet  (\ref{yyy3.22}) and  (\ref{yyy3.23}) of $o(2,1)$  realizations
is possible, because the transformation $H \to -H$,
 $K \to - K$, $D\to D$ leaves the subalgebra (\ref{su11}) invariant.

 In order to fulfill the covariance relations (\ref{H-A}--\ref{D-A}) we should choose the following matrix realizations
 \begin{eqnarray}\label{yyy3.24}
 P_i &=& \sigma_i \otimes \sigma_{-} \otimes {\mathscr{C}}_P\,,
 \qquad {\mathscr{C}}_P = p_0 + \vec{p}\,\vec{\sigma} \,,
 \cr
  F_i &=& \sigma_i \otimes \sigma_{+} \otimes {\mathscr{C}}_F\,,
 \qquad {\mathscr{C}}_F = f_0 + \vec{f}\,\vec{\sigma} \,.
 \end{eqnarray}
 Using two possible definitions  of $B_i$ - by commutators $[K, P_i]$ and $[H,F_i]$ - one obtains that
 \begin{equation}\label{yyy3.25}
 p_0 = f_0\,, \qquad p_1 = - f_1 \,,
 \qquad p_2 = - f_2 \,,\qquad p_3 = f_3\,,
 \end{equation}
 and
 \begin{equation}\label{yyy3.26}
 B_i  = {\textstyle\frac{i}{2}}\,\sigma_i \otimes
 \Big(
 \sigma_3 \otimes (p_3 + p_0 \sigma_3)  - \mathbf{1}_2 \otimes \left( i \, \sigma_1 p_2 - i\, \sigma_2 p_1 \right)
 \Big)\,.
 \end{equation}
 Now we shall consider the relations (\ref{A-A-com1}).
 The commutators $[P_i, P_j]= [F_i, F_j]=0$ are satisfied for any choice of ${\mathscr{C}}_P$,  but from
 $[P_i, F_j]=0$ follows that
 \begin{equation}\label{yyy3.27}
 {\mathscr{C}}_P \, {\mathscr{C}}_F = 0 \qquad \Rightarrow \qquad p^2_0 - p^2_1 - p^2_2 = 0\,,  \quad p_3 =0 \,.
 \end{equation}
 It can be checked that the Abelian commutativity conditions involving $B_i$ are satisfied if the conditions (\ref{yyy3.27}) are valid.

The matrix realizations (\ref{yyy3.24}) lead to Hermitian quantum-mechanical generators $\hat{P}_i, \hat{F}_i$ (see (\ref{real-op1}))
if the matrices ${\mathscr{C}}_P$ and ${\mathscr{C}}_F$ are Hermitian, what implies that
$p_0, p_1, p_2$ should be real.
But, in opposite to the ``Euclidean'' case (\ref{yyy3.13}), the quantities  $(p_0, p_1, p_2)$
form isotropic Minkowski three--vector.
Thus, the nonzero solutions of the real constraints (\ref{yyy3.27}) exist (e.g. $p_0{=}p_1$, $p_2{=}p_3{=}0$), i.e. the
corresponding Hermitian generators of GCA also exist.

 The argument can be extended to $N$-twistorial system defining the following indefinite Hamiltonian
 \begin{equation}\label{yyy3.28}
 \hat H = \sum\limits^{N}_{k=1} d_k \, \bar t^{(k)}_{\dot{\alpha},1} \, t^{(k)}_{\alpha,1} \,, \qquad
 d_k= \pm 1 \,.
 \end{equation}
 Introducing the matrix $\Delta_N = \hbox{diag} (d_1 \ldots , d_N)$ the matrix realization
 which leads via (\ref{real-op1}) to (\ref{yyy3.28}) takes the form
 \begin{equation}\label{yyy3.29}
 H =  \mathbf{1}_2 \otimes \sigma_{-} \otimes \Delta_N\,.
 \end{equation}
 From (\ref{su11}) uniquely follows the choice
 \begin{eqnarray}\label{yyy3.30}
 K & {=} &  \mathbf{1}_2 \otimes \sigma_{+} \otimes \Delta_N \,,
 \cr
 D & {=} &  \mathbf{1}_2 \otimes {\textstyle\frac{i}{2}} \, \sigma_3\otimes \mathbf{1}_N \,.
 \end{eqnarray}
 We put further
 \begin{equation}\label{yyy3.31}
 M_i = {\textstyle\frac{1}{2}} \, \sigma_i \otimes \mathbf{1}_2 \otimes \mathbf{1}_{2N}\,.
 \end{equation}
 If we postulate that
 \begin{equation}\label{yyy3.32}
 P_i = \sigma_i \otimes P\,, \qquad
 B_i = \sigma_i \otimes B\,, \qquad
 F_i=\sigma_i \otimes F \,,
 \end{equation}
 where $P$, $B$  and $F$ are the $2N{\times}2N$ matrices, from covariance properties (\ref{yyy2.18}) one gets the formulae  (${\mathscr{F}}$ is an $N{\times}N$ arbitrary matrix)
 \begin{eqnarray}
 P & = &  \sigma_{-} \otimes \Delta_N \, {\mathscr{F}} \, \Delta_N
  \,,
 \nonumber\\
 F & = &  \sigma_{+} \otimes {\mathscr{F}} \,,
 \label{yyy3.33}\\
 B & = & \textstyle{\frac{i}{2}}\,
 \Big(
 \mathbf{1}_2 \otimes [\Delta_N, {\mathscr{F}}]
 - \sigma_3 \otimes \{ \Delta_N, {\mathscr{F}} \}
 \Big)  \,. \nonumber
 \end{eqnarray}
 The Abelian structure of subalgebra ($P_k, B_k, F_k$) leads to the  following condition
 \begin{equation}\label{yyy3.34}
\left(
\Delta_N \,{\mathscr{F}}
\right)^2 = 0 \quad  \leftrightarrow
\quad
  {\mathscr{F}} \, \Delta_N \,
 {\mathscr{F}} = 0 \,,
\end{equation}
which is equivalent to $({\mathscr{F}} \Delta_N)^2=0$.
 Further
Hermiticity of quantum--mechanical generator $\hat{P}_i$ implies that $\Delta_N {\mathscr{F}} \Delta_N$ is Hermitian what is equivalent to the Hermiticity of matrix ${\mathscr{F}}$.
We see therefore that Hermitian realizations  require nonvanishing  solutions of (\ref{yyy3.34}) with Hermitian ${\mathscr{F}}$.
For any $\Delta_N{\neq}\mathbf{1}_{N}$ such nonvanishing ${\mathscr{F}}$ exists.
For example, in the case $\Delta_N=\hbox{diag} (\mathbf{1}_p , -\mathbf{1}_q)$, $p+q=N$, $p>q$, the general solution
for matrix ${\mathscr{F}}$ is the following
 \begin{equation}\label{matrF}
{\mathscr{F}}=
\left(
  \begin{array}{ccc}
    \mathbf{0}_{p-q} & \mathbf{0} & \mathbf{0} \\
    \mathbf{0} & M & MU \\
    \mathbf{0} & MU^+ & M \\
  \end{array}
\right)
 \,,
\end{equation}
where $q{\times}q$ matrices $M$ and $U$ commute, $[M,U]=0$, $M$ is Hermitian ($M=M^+$),
and $U$ is unitary ($UU^+=1$).

We see therefore that if $\Delta_N{\neq}\mathbf{1}_{N}$, any twistor realization (\ref{yyy3.28})
of the Hamiltonian generator can be extended to the realization of complete GCA.

\subsection{Hermitian twistorial realization of Galilean conformal algebra in $N{=}\,2$ case}

\quad\, Finally we shall present explicit $N{=}\,2$ Hermitian representation of GCA.
Taking matrix realization  (\ref{yyy3.22})-(\ref{yyy3.24}), (\ref{yyy3.26}) with $p_0{=}p_1{=}\frac12$, $p_2{=}p_3{=}0$
and using basic construction (\ref{real-op1}) we obtain the following quantum--mechanical twistorial realization of GCA
\begin{equation}\label{M-fin}
\hat M_i= {\textstyle\frac{1}{2}} \sum_{k=1}^2 \left(\bar t_{\dot\alpha ,1}^{(k)}(\sigma_i)_{\alpha}{}^{\beta}t_{\beta ,2}^{(k)}-
\bar t_{\dot\alpha ,2}^{(k)}(\sigma_i)_{\alpha}{}^{\beta}t_{\beta ,1}^{(k)}\right)\,,
\end{equation}
\begin{eqnarray}
\hat H&=& \bar t_{\dot{\alpha}, 1}^{(1)}\, t_{\alpha, 1}^{(1)} - \bar t_{\dot{\alpha}, 1}^{(2)}\, t_{\alpha, 1}^{(2)}\,, \nonumber \\
\hat K&=& \bar t_{\dot{\alpha}, 2}^{(1)}\, t_{\alpha, 2}^{(1)}- \bar t_{\dot{\alpha}, 2}^{(2)}\, t_{\alpha, 2}^{(2)}\,, \label{R-fin}\\
\hat D&=&{\textstyle\frac{i}{2}}\sum\limits_{k=1}^2 \left(\bar t_{\dot\alpha ,2}^{(k)}\,t_{\alpha ,1}^{(k)}-
\bar t_{\dot\alpha ,1}^{(k)}\,t_{\alpha ,2}^{(k)}\right) \nonumber
\end{eqnarray}
\begin{eqnarray}
\hat P_i&=& \bar t_{\dot\alpha , 1}^{(1)} \,(\sigma_i)_{\alpha}{}^{\beta}\, t_{{\beta},1}^{(1)}+
\bar t_{\dot\alpha , 1}^{(2)} \,(\sigma_i)_{\alpha}{}^{\beta}\, t_{{\beta},1}^{(2)}+
\bar t_{\dot\alpha , 1}^{(1)} \,(\sigma_i)_{\alpha}{}^{\beta}\, t_{{\beta},1}^{(2)}+
\bar t_{\dot\alpha , 1}^{(2)} \,(\sigma_i)_{\alpha}{}^{\beta}\, t_{{\beta},1}^{(1)}\,, \nonumber \\
\hat F_i &=& \bar t_{\dot\alpha , 2}^{(1)}\,(\sigma_i)_{\alpha}{}^{\beta}\,t_{{\beta},2}^{(1)}+
\bar t_{\dot\alpha , 2}^{(2)}\,(\sigma_i)_{\alpha}{}^{\beta}\,t_{{\beta},2}^{(2)}-
\bar t_{\dot\alpha , 2}^{(1)}\,(\sigma_i)_{\alpha}{}^{\beta}\,t_{{\beta},2}^{(2)}-
\bar t_{\dot\alpha , 2}^{(2)}\,(\sigma_i)_{\alpha}{}^{\beta}\,t_{{\beta},2}^{(1)}\,, \label{A-fin}\\
\hat B_i&=&{\textstyle\frac{i}{2}}\left(\bar t_{\dot\alpha ,2}^{(1)}\,(\sigma_i)_{\alpha}{}^{\beta}\,t_{\beta ,1}^{(1)}-
\bar t_{\dot\alpha ,1}^{(1)}\,(\sigma_i)_{\alpha}{}^{\beta}\,t_{\beta ,2}^{(1)}-
\bar t_{\dot\alpha ,2}^{(2)}\,(\sigma_i)_{\alpha}{}^{\beta}\,t_{\beta ,1}^{(2)}+
\bar t_{\dot\alpha ,1}^{(2)}\,(\sigma_i)_{\alpha}{}^{\beta}\,t_{\beta ,2}^{(2)}\right) \nonumber \\
&&+{\textstyle\frac{i}{2}}\left(\bar t_{\dot\alpha ,2}^{(1)}\,(\sigma_i)_{\alpha}{}^{\beta}\,t_{\beta ,1}^{(2)}+
\bar t_{\dot\alpha ,1}^{(1)}\,(\sigma_i)_{\alpha}{}^{\beta}\,t_{\beta ,2}^{(2)}-
\bar t_{\dot\alpha ,2}^{(2)}\,(\sigma_i)_{\alpha}{}^{\beta}\,t_{\beta ,1}^{(1)}-
\bar t_{\dot\alpha ,1}^{(2)}\,(\sigma_i)_{\alpha}{}^{\beta}\,t_{\beta ,2}^{(1)}\right)\,. \nonumber
\end{eqnarray}

From  (\ref{A-fin}) follows apparently that the operators $\hat P_i$, $\hat F_i$ and $\hat B_i$ are Hermitian.
Their commutativity follows if we note, that after introducing of the operators
\begin{equation}\label{new-op}
u_{\alpha}\equiv t_{{\alpha},1}^{(1)}+\epsilon_{\alpha\beta}\,\bar t_{\dot\beta , 1}^{(2)}\,,\qquad
v_{\alpha}\equiv t_{{\alpha},2}^{(1)}-\epsilon_{\alpha\beta}\,\bar t_{\dot\beta , 2}^{(2)}\,,
\end{equation}
where $\epsilon_{\alpha\beta}=-\epsilon_{\beta\alpha}$, $\epsilon_{12}=1$ is a second ${\rm SU}(2)$--invariant
tensor in addition to $\delta_{\alpha\dot\beta}$ (see footnote 4),
the operators forming the Abelian subalgebra (\ref{A-fin}) can be represented as follows
\begin{eqnarray}
\hat P_i&=& \bar u_{\dot\alpha} \,(\sigma_i)_{\alpha}{}^{\beta}\, u_{{\beta}}\,, \nonumber \\
\hat F_i &=& \bar v_{\dot\alpha} \,(\sigma_i)_{\alpha}{}^{\beta}\, v_{{\beta}} \,, \label{A1-fin}\\
\hat B_i&=&{\textstyle\frac{i}{2}}\Big(\bar v_{\dot\alpha} \,(\sigma_i)_{\alpha}{}^{\beta}\, u_{{\beta}}-
\bar u_{\dot\alpha} \,(\sigma_i)_{\alpha}{}^{\beta}\, v_{{\beta}}\Big) \,. \nonumber
\end{eqnarray}
We observe that the operators  (\ref{new-op})  and its conjugates describe only half of the $N{=}\,2$ twistorial degrees of freedom
and it follows from (\ref{CCR-tw1}) that they commute between themselves,
\begin{equation}\label{com-new}
[u_{\alpha},u_{\beta}]=[u_{\alpha},\bar u_{\dot\beta}]=[u_{\alpha},v_{\beta}]=
[u_{\alpha},\bar v_{\dot\beta}]=[v_{\alpha},\bar v_{\dot\beta}]=[v_{\alpha},v_{\beta}]=0\,.
\end{equation}
As result, we see from the formulae (\ref{A1-fin}) and (\ref{com-new}) that our realization
for $\hat P_i$, $\hat F_i$, $\hat B_i$ forms a nine--dimensional Abelian subalgebra.

\section{Final Remarks}

\quad\, An important advantage of the relativistic Penrose twistor formulation is the possibility of expressing the generators of
${\rm O}(4,2)$ conformal algebra as current--like bilinear expressions in $D{=}\,4$ quantized twistor variables.
Unfortunately in nonrelativistic case the GCA generators
can be expressed bilinearly by quantized Galilean twistors only if the Hamiltonian is not positive--definite.
We recall that the difference between relativistic and Galilean conformal groups is essential:
the relativistic one is semisimple, and the nonrelativistic is described by a semidirect product
involving for $d{=}\,3$ nine--dimensional Abelian subalgebra. It should be stressed that however the GCA is
the $c{\to}\,\infty$ contraction limit of relativistic $o(4,2)$ algebra,
we were not able to obtain the Galilean $N$--twistor realizations by the contraction  $c{\to}\,\infty$
of the relativistic $N$--twistor realizations. Similarly, if we introduce explicitly the nonrelativistic
space--time by rescaling $x_0{=}\,ct$, one does not get the nonrelativistic limit of the
incidence relations  (\ref{inci-tw}). The problem of expressing nonrelativistic space--time
$(x_i, t)$ as composite in terms of Galilean twistors requires further investigation.

Finally we recall that the nonrelativistic GCA has been recently extended to $n$--extended
Galilean superconformal algebra \cite{AzLu,Saka} with $n{=}\,2k$. E.g. for the lowest--dimensional
case $n{=}\,2$ it would be interesting to prove the expected result that using the pair
of $n{=}2$ nonrelativistic supertwistors one can construct a Hermitian supertwistorial
realization of Galilean  $n{=}\,2$ superconformal algebra but it will lead as well to indefinite
generator $H$.

In this short Letter, we do not discuss the application of our considerations to concrete models
with indefinite Hamiltonian.
We only note here that indefinite Hamiltonian was also obtained in the quantum--mechanical model with
$D{=}\,2{+}1$ ``exotic'' Galilean symmetry \cite{LSZ97},
but the ghosts in \cite{LSZ97} were eliminated by imposing the additional condition.
Certain subsidiary conditions  excluding the states with negative energy were also considered in some models
of supersymmetric mechanics and field theory \cite{SSV}.

\bigskip\bigskip
\noindent {\bf\large Acknowledgements}

\noindent
We would like to thank Evgeny Ivanov for his interest in the Letter and valuable comments.
S.F. thanks also Andrei Smilga for the discussion on the subject of this Letter.
We acknowledge support from a grant of the Bogoliubov-Infeld Programme, RFBR
grants 09-02-01209, 09-01-93107 (S.F.) and
from Polish Ministry of Science and Higher Educations, grant No.~N202331139 (P.K., J.L. and P.M.).
S.F. would like to thank
the Institute for Theoretical Physics at Wroclaw University
for its warm hospitality at the first stage of this study.

\end{document}